\documentclass[onecolumn,prl,showpacs,multicol,amsmath,amssymb]{revtex4-2}
\usepackage{amssymb}
\usepackage{amsmath}
\usepackage{graphicx}
\usepackage{color}
\usepackage{epstopdf}
\usepackage{bm}
\usepackage{gensymb}

\newcommand\redout{\bgroup\markoverwith{\textcolor{red}{\rule[.5ex]{2pt}{0.4pt}}}\ULon}
\usepackage[colorlinks=true,citecolor=blue]{hyperref}
\hypersetup{colorlinks=true,citecolor=blue,linkcolor=red,urlcolor=blue}

\graphicspath{{.}{./plot/}}

\begin{document}
	
	\title{Atomic Interferometer Gates Realization via Quantum Optimal Control}
	
	\author{Javad Sharifi}\email{jv.sharifi@gmail.com}
	\affiliation{Electrical and Computer Engineering Department, Qom University of Technolgy, Qom, Iran}
	
	\date{\today}
	
	\begin{abstract}  
		Atomic Interferometer has two quantum unitary gates that must be realized for quantum sensing purposes from atomic gravimeter and atomic interferometer gyroscope. An optimal cost function which define the distance between two unitary operator is defines and based on it a pulse based stochastic gradient descent algorithm is derived for implementation of atomic mirror and atomic beam-splitter gates. By using numerical optimal control for those gate realization, we achieved to gate fidelity of 0.995 and 0.998 for mirror and beam-splitter gates, respectively. For this research, a two-level atomic clock transitions are employed.
	\end{abstract}
	
	\maketitle
	
	\section{Introduction}
	Atoms are surprisingly accurate measurement devices. They can detect external signals with high sensitivity furthermore of very accurate time measurement in atomic clocks.  Important class of atom sensors are atomic interferometers that in a controlled manner have versatilized to investigate fundamental physics such as particle physics, general relativity\cite{chou2010optical} , gravity\cite{wu2019gravity,kovachy2015quantum}, gravitational wave detection\cite{loeb2015using,abe2021matter}, dark matter \cite{arvanitaki2015searching}  and cosmology. It also has many applications in quantum sensing application and quantum metrology, navigation and geophysics.
	Zhaoshan long-baseline atom interferometer gravitation antenna (ZAIGA) in China\cite{zhan2020zaiga}, 100-meter Magis-100 atom interferometer in Fermilab in USA\cite{abe2021matter}
	For detailed formulation of atomic interferometry see the lecture\cite{kovachymacroscopic}.  Atomic Gyroscopes are hoped be more sensitive with respect to optical gyroscopes and near future to mechanical gyroscopes. Atom interferometers performs sensitivity better than light interferometers,  because the de Broglie wavelength is more than several orders of magnitude smaller than the optical wavelength. Hence, for atom and light gyroscopes of the same area, small displacements cause much larger phase shifts for atoms and consequently better sensitivity based on Sagnac phase shift formula\cite{fang2012advances, gustavson2000precision}. 
	
	In otherwise, the precise working of any technology depends on its embedded control system, and especially, universal quantum computation requires the implementation of arbitrary control operations on the quantum register\cite{zhang2015experimental}. Among different science and technologies, quantum science and technology will become dominant future technology, which will change the whole life of human beings. For quantum technology, it is essential to develop quantum materials, devices, and circuits then precisely engineer and control quantum states or unitary gates of such devices based on appropriate high fidelity circuit readout.
	For example, for quantum computing application, exact unitary matrix of quantum gates must be implemented in spite of decoherence, relaxation, and in some quantum, materials mitigate leakage errors which lead to exit from quantum computational space. All of this issues furthermore of development and engineering quantum designs needs precise and high-fidelity quantum control protocol \cite{kwon2021gate,krantz2019quantum,haffner2008quantum,weiss2017quantum,henriet2020quantum} to be implemented on appropriate classical hardware such as SOC-FPGA and quantum hardware. 
	Quantum control as essential part of development of quantum science and technology dates back to 1980-2000 by pioneering research works in\cite{huang1983controllability,belavkin1983theory,warren1993coherent, chu2002cold}.
	 optimal control especially numerical pulse based optimal control does perform better than  quantum control methods since they guarantee optimal of realization and speed of convergence to optimal gates or states.
	 
	 In this research, at first we introduce the quantum dynamics of two-level atomic transition clock and also atomic interferometer, then we introduce the objective cost for gate realization and derive the numerical pulse based optimal control for this quantum system. Finally numerically simulate the optimal control for implementation of two unitary gates of atomic interferometer, i.e. reflection or mirror and beam-splitter gate.
	
	\begin{figure}
		\centerline{\includegraphics[width=1\columnwidth]{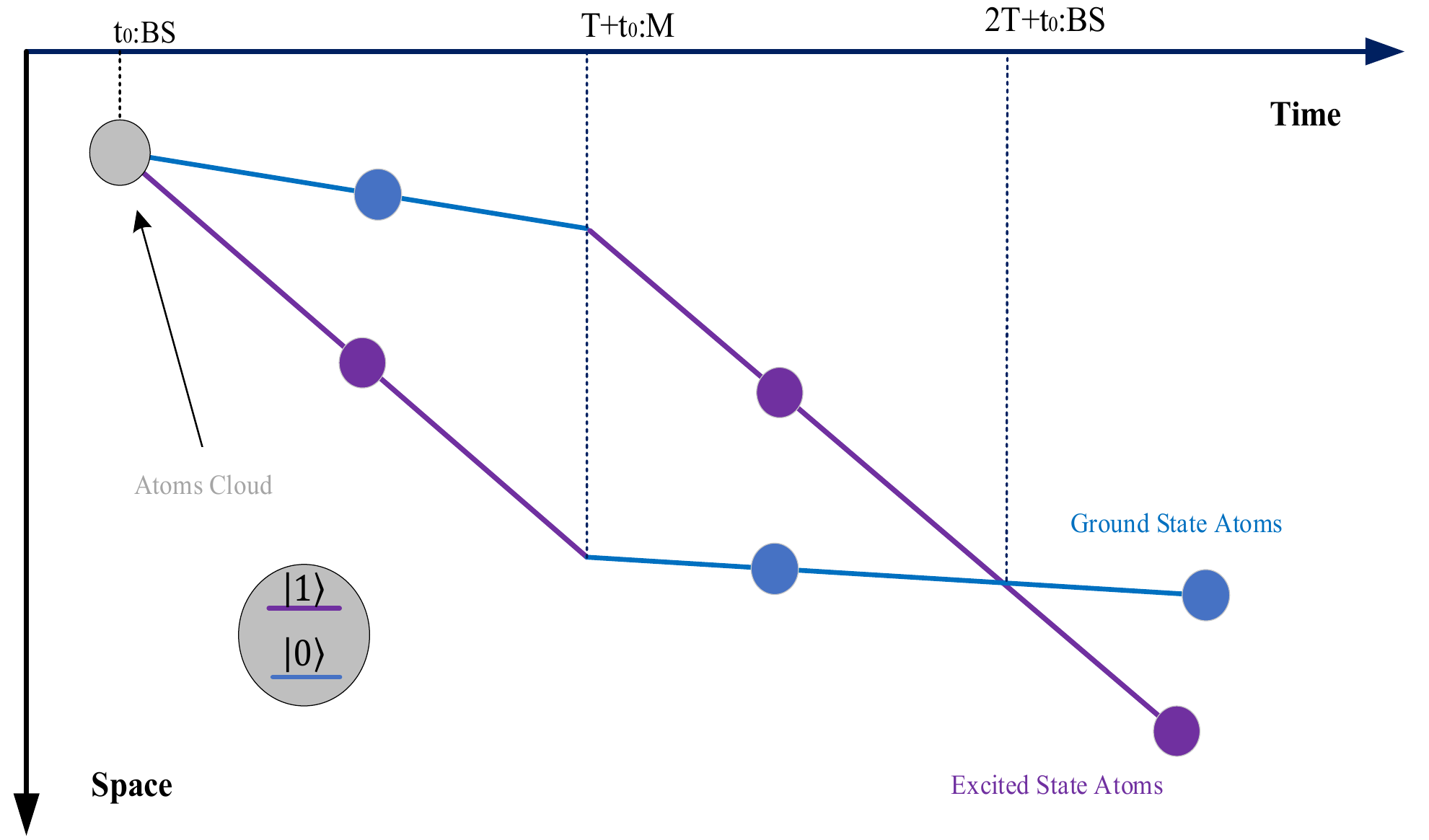}}
		\caption{Mach-Zehnder based Atom Interferometer in space-time coordinate: At times $t_0,2T+t_0$ the atomic beam splitter (BS) and at time $T+t_0$ the atomic mirror (M) must be implemented.}
		\label{1}
	\end{figure}
	
	\section{Two-Level Atom Rabi Oscillation as Rotation over Bloch Sphere}
	These quantum mechanical matter waves can be manipulated with the atomic equivalents of lenses, beam-splitters, and mirrors. Atom interferometry is analogous to optical interferometry. In both interferometers, an incident wave is split into two paths by a beam-splitter. The two paths are later redirected back toward each other with mirrors and overlapped on a final beam-splitter to produce an interference pattern. It is known how to make beam-splitter and mirror for optical beams. To implement the similar interferometer devices, in experiment, it will considered an ensemble of cold two-level atoms such as  clock transition states ($|0\rangle={}^1 \mathrm{S}_0 \leftrightarrow  |1\rangle={}^3\mathrm{P}_0 $) of Strontium atom (${}^{87}\mathrm{Sr}$) with a $\lambda_{|0\rangle \leftrightarrow |1\rangle}=698$nm wavelength transition and very long coherence time. 
	Freely falling atoms have enabled the development of atomic gravimeters (Kasevich and Chu, 1992; Peters, Chung, and Chu, 1999) and gyrometers (Gustavson, Bouyer, and Kasevich, 1997; Gustavson, Landragin, and Kasevich, 2000). In these devices an atomic cloud measures acceleration by sensing the spatial phase shift of a laser beam along its freely falling trajectory.
	
	The quantum system described by Schrödinger equation as $i\hslash\frac{d}{dt}|\psi_
	t\rangle=H(t)|\psi_t\rangle$ and its solution is $|\psi_t\rangle=e^{\frac{-i}{\hslash}\int_0^t H(t)dt}|\psi_0\rangle=U(t) |\psi_0\rangle$, for very small time interval $dt=\delta t\rightarrow 0$, the integral can be break to sum of $k$ intergral as $\int_0^t H(t)dt=\sum_{n=0}^{k} \int_{n\delta t}^{(n+1)\delta t} H(n\delta t)dt=\sum_{n=0}^k H(n\delta t)\delta t$, then we have $U(t)=e^{\frac{-i}{\hslash} \sum_{n=0}^{k}H(n\delta t)\delta t}$. The Hamiltonian of two-level atom in Rabi modelling is:
	\begin{align}
		\label{two-level}
		H(t)=\frac{\hbar}{2}	
		\begin{pmatrix}
			\Delta(t)&\Omega(t)e^{-i\phi}\\\Omega(t)e^{i\phi}&-\Delta(t)
		\end{pmatrix}
	\end{align}
	in which $\Omega , \Delta$ are Rabi frequency and Detuning respectively and $\phi$ is the initial phase of laser electric field. Sometimes scientist employ the three-level Raman transition atoms for atom interferometer\cite{....}, in this case there are two stable atomic levels $|0\rangle , |1\rangle$ and unstabe excited state $|e\rangle$. For Raman three levels atoms there are two laser beams, one is the upward beam with Rabi frequency $\Omega_1$ and phase $\phi_1$ which transfer the atomic state transition $|0\rangle \rightarrow |e\rangle$ and the other is downward beam with Rabi frequency $\Omega_2$ and phase $\phi_2$ which lead to transition $|e\rangle \rightarrow |1\rangle$, the detuning in both transitions assumes to be approximately equal to $\Delta$. It is shown in\cite{berman1997atom} that for $\Delta \gg \Omega_1 , \Omega_2$, the three-level Raman transition atom approximate with two-level atom with $\Omega=\frac{\Omega_1 \Omega_2}{2\Delta} , \phi=\phi_2-\phi_1$, then for every interferometer we can use two-level Hamiltonian model of equation\ref{two-level}. We can write this Hamiltonian based on Pauli matrices  $\sigma_{x},\sigma_{y},\sigma_{z}$ as follows:
	\begin{align}
		H(t)=\frac{\hbar}{2}\Big(\Omega(t)\mathrm{cos}(\phi)\sigma_{x}+\Omega(t)\mathrm{sin}(\phi)\sigma_{y}+\Delta(t)\sigma_{z}\Big)
	\end{align}
	Then the unitary evolution of two-level atom at time sample $t=k\delta t$ is:
	
	\begin{equation}
		\label{unitary}
		U(k\delta t)=exp\Big(-\frac{i}{2}\delta t(\mathrm{cos}(\phi)\sigma_{x}+\mathrm{sin}(\phi)\sigma_{y})\sum_{n=0}^k\Omega (n\delta t)-\frac{i}{2}\delta t\sigma_{z}\sum_{n=0}^k \Delta(n\delta t)\Big)
	\end{equation}
	
	We now that, the unitary rotation operator for transition of quantum states is\cite{glendinning2010rotations}:
	\begin{align}
		\label{rotation}
		U(t)=e^{-i\frac{\alpha(t)}{2}\hat{r}(t).\vec{\sigma}}=\mathrm{sin}(\frac{_1}{^2}\alpha(t))I-i\mathrm{cos}(\frac{_1}{^2}\alpha(t))\hat{r}(t).\vec{\sigma}
	\end{align}
	with $\vec{\sigma}=\sigma_{x}\hat{x}+\sigma_{y}\hat{y}+\sigma_{z}\hat{z}$  and $\hat{r}=r_{x}\hat{x}+r_{y}\hat{y}+r_{z}\hat{z}$ is the rotation axis and $\alpha=\omega_{q} t$ is rotation angle, $\omega_{q}$ is the angular speed of quantum state vector on Bloch sphere. By comparing equations \eqref{unitary},\eqref{rotation} , we will obtain:
	\begin{align}
		\alpha \hat{r}=\delta t\Big( \mathrm{cos}(\phi)\sum_{n=0}^k \Omega(n\delta t)\hat{x}+\mathrm{sin}(\phi)\sum_{n=0}^k \Omega(n\delta t)\hat{y}+\sum_{n=0}^k \Delta(n\delta t)\hat{z}\Big)
	\end{align}
	then since $|\alpha \hat{r}|=\alpha$ we obtain:
	
	\begin{align}
		\alpha(k\delta t)=\delta t\sqrt{(\sum_{n=0}^k \Omega(n\delta t))^2+(\sum_{n=0}^k \Delta(n\delta t))^2}=\delta t\sqrt{(\sum_{n=0}^k \Omega_n)^2+(\sum_{n=0}^k \Delta_n)^2}
	\end{align}
	
	\begin{align}
		\hat{r}(k\delta t)=\mathrm{cos}(\phi)\frac{\sum_{n=0}^k \Omega_n}{\sqrt{(\sum_{n=0}^k \Omega_n)^2+(\sum_{n=0}^k \Delta_n)^2}}\hat{x}+\mathrm{sin}(\phi)\frac{\sum_{n=0}^k \Omega_n}{\sqrt{(\sum_{n=0}^k \Omega_n)^2+(\sum_{n=0}^k \Delta_n)^2}}\hat{y}+\frac{\sum_{n=0}^k \Delta_n}{\sqrt{(\sum_{n=0}^k \Omega_n)^2+(\sum_{n=0}^k \Delta_n)^2}}\hat{z}
	\end{align}
	in these relations, we use abbreviates $\Delta_n=\Delta(n\delta t), \Omega_n=\Omega(n\delta t)$.
	
	\section{Optimal Control Implementation of Mirror and Beam-Splitter Gates}
	The unitary operator of a mirror (M) and beam-splitter (BS) is:
	\begin{align}
		\label{gates}
		U_{M}=\begin{pmatrix}
			0&&1\\
			1&&0
		\end{pmatrix}
		=\sigma_{x} , U_{BS}=\frac{1}{\sqrt{2}}\begin{pmatrix}
			1&&i\\
			i&&1
		\end{pmatrix}
	\end{align}
	
	The aim of gate optimization is the implementation of atomic unitary operator of equation \eqref{unitary} goes to desired gate of equation \eqref{gates}. For this aim, from quantum computing community, we can select an objective function as:
	\begin{align}
		J_{G}=1-\frac{1}{d^2}\Big|\mathrm{Tr}\big(U_{G}^\dag U_{atom}\big)\Big|^2
	\end{align}
	in which $U_{atom}=U(k\delta t), G=M,BS, d=\mathrm{Tr}(U_{G}^{\dag}U_{G})=2$. For this cost function as $U_{atom} \rightarrow U_{G}$ the $J_{gate} \rightarrow 0$. The numerical optimal contro with stochastic gradient descent (SGD), would iterate the Rabi frequency and detuning as follows: 
	\begin{align}
		\begin{bmatrix}
			\Omega_{k+1}\\
			\Delta_{k+1}
		\end{bmatrix}=
		\begin{bmatrix}
			\Omega_{k}\\\Delta_{k}
		\end{bmatrix}
		-
		\begin{bmatrix}
			\eta_{\Omega}&&0\\
			0&&\eta_{\Delta}
		\end{bmatrix}
		\begin{bmatrix}
			\frac{\partial J_{gate}}{\partial\Omega_{k}}\\
			\frac{\partial J_{gate}}{\partial \Delta_{k}}
		\end{bmatrix}
	\end{align}
	
	At first we choose to implement the mirror gates, then we will obtain the following cost function:
	\begin{align}
		J_{M}=1-\mathrm{cos}^2(\frac{_1}{^2}\alpha)r^2_{x}=1-\mathrm{cos}^2(\phi)\mathrm{cos}^2\Big(\frac{\delta t}{2}\sqrt{(\sum_{n=0}^k \Omega_n)^2+(\sum_{n=0}^k \Delta_n)^2}\Big)\frac{(\sum^k_{n=0}\Omega_{n})^2}{(\sum^k_{n=0}\Omega_{n})^2+(\sum^k_{n=0}\Delta_{n})^2}
	\end{align}
	then, the following updates will obtained:
	\begin{align}
		\frac{\partial J_{M}}{\partial \Omega_{k}}=
		\frac{1}{2}\delta t \mathrm{cos}^2(\phi)\mathrm{sin}(\alpha)\frac{(\sum_0^k \Omega_{n})^3}{\Big((\sum^k_{n=0}\Omega_{n})^2+(\sum^k_{n=0}\Delta_{n})^2\Big)^{\frac{3}{2}}}-2\mathrm{cos}^2(\phi)\mathrm{cos}^2(\frac{\alpha}{2})\frac{(\sum_0^k \Omega_{n})(\sum_0^k \Delta_{n})^2}{\Big((\sum^k_{n=0}\Omega_{n})^2+(\sum^k_{n=0}\Delta_{n})^2\Big)^2}
	\end{align}
	
	\begin{align}
		\frac{\partial J_{M}}{\partial \Delta_{k}}=
		\frac{1}{2}\delta t \mathrm{cos}^2(\phi)\mathrm{sin}(\alpha)\frac{(\sum_0^k \Delta_{n})(\sum_0^k \Omega_{n})^2}{\Big((\sum^k_{n=0}\Omega_{n})^2+(\sum^k_{n=0}\Delta_{n})^2\Big)^{\frac{3}{2}}}+2\mathrm{cos}^2(\phi)\mathrm{cos}^2(\frac{\alpha}{2})\frac{(\sum_0^k \Delta_{n})^3}{\Big((\sum^k_{n=0}\Omega_{n})^2+(\sum^k_{n=0}\Delta_{n})^2\Big)^2}
	\end{align}
	
\begin{figure}
	\centerline{\includegraphics[width=1\columnwidth]{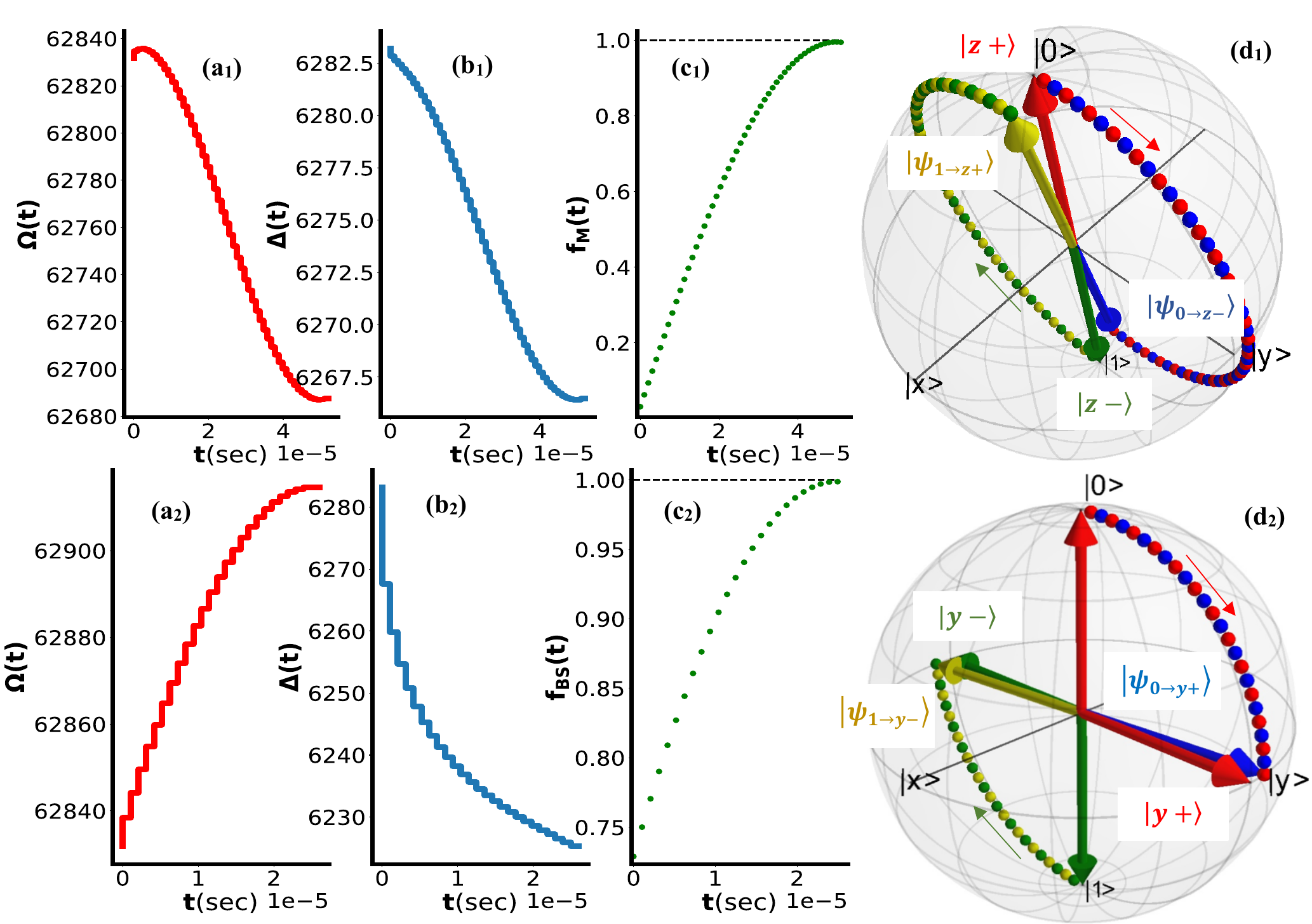}}
	\caption{($a_1,b_1$) are the control pulses and ($c_1$) fidelity function for implementation of atomic mirror gate realization and ($d_1$) shows the simultaneous quantum state transfer ($|0\rangle \rightarrow |1\rangle$ and $|1\rangle \rightarrow |0\rangle$) on Bloch-sphere  using optimal reflection gates realization. Figures ($a_2,b_2,c_2$) are the control pulses and fidelity over time for beam-splitter implementation. ($d_2$) displays simultaneous quantum state transfer ($|0\rangle \rightarrow \frac{1}{\sqrt{2}}(|0\rangle+i|1\rangle)$ and $\frac{1}{\sqrt{2}}(i|0\rangle+|1\rangle) \rightarrow |1\rangle$) using optimal beam-splitter gates realization on Bloch-sphere. }
	\label{2}
	\end{figure}

The Optimization be possible, we must have the initial phase of Rabi signal as $\phi \neq\frac{(2n+1)\pi}{2}$.  The control signals are depicted in figure\ref{2}($\mathrm{a_1,b_1}$) and the fidelity of gate optimization is depicted in figure\ref{2}($\mathrm{c_1}$). After 50 sample of 1 microsecond control pulses we reach to the fidelity of $0.995$ for mirror gate optimization and the atomic gate at this optimal control reach to:
	\begin{align}
		U_{atomM}(N\delta t)=\begin{pmatrix}
			0.0239-0.0994i&&0.9947i\\
			0.9947i&&0.0239+0.0994i
		\end{pmatrix} , N=50 , \delta t=1\mu sec
	\end{align}
	By gate optimization, simultaneously we can control two populations of atoms, one population from state $|0\rangle$ to $|1\rangle$ and other population from $|1\rangle$ to $|0\rangle$ as we can see in figure\ref{2}($\mathrm{d_1}$).

	Now, to optimally implement the atomic beam-splitter on atomic clouds, we will optimize the following cost function:
	\begin{align}
		J_{BS}=1-\frac{1}{2}\Big(\mathrm{sin}(\frac{\alpha}{2})-\mathrm{cos}(\frac{\alpha}{2})r_{x}\Big)^2
	\end{align}
	
	and after simple calculations, we will have the following update rules for Rabi frequency and detunings:
	\begin{align}
		\frac{\partial J_{BS}}{\partial \Omega_{k}}=-\Big( \mathrm{sin}(\frac{\alpha}{2})-\mathrm{cos}(\frac{\alpha}{2})r_{x}
		\Big)
		\Big((\frac{\delta t}{2}\mathrm{cos}(\frac{\alpha}{2})+\frac{\delta t}{2}\mathrm{sin}(\frac{\alpha}{2})r_{x})\frac{\sum_0^k \Omega_{n}}{\sqrt{(\sum^k_{n=0}\Omega_{n})^2+(\sum^k_{n=0}\Delta_{n})^2}}-\mathrm{cos}(\frac{\alpha}{2})\mathrm{cos}(\phi)\times\\ \nonumber
		\frac{(\sum_0^k \Delta_{n})^2}{\Big((\sum^k_{n=0}\Omega_{n})^2+(\sum^k_{n=0}\Delta_{n})^2\Big)^{\frac{3}{2}}}\Big)
	\end{align}
	
	\begin{align}
		\frac{\partial J_{BS}}{\partial \Delta_{k}}=-\Big( \mathrm{sin}(\frac{\alpha}{2})-\mathrm{cos}(\frac{\alpha}{2})r_{x}
		\Big)
		\Big((\frac{\delta t}{2}\mathrm{cos}(\frac{\alpha}{2})+\frac{\delta t}{2}\mathrm{sin}(\frac{\alpha}{2})r_{x})\frac{\sum_0^k \Delta_{n}}{\sqrt{(\sum^k_{n=0}\Omega_{n})^2+(\sum^k_{n=0}\Delta_{n})^2}}+\mathrm{cos}(\frac{\alpha}{2})\mathrm{cos}(\phi)\times\\ \nonumber
		\frac{\sum_0^k \Delta_{n}\sum_0^k\Omega_n}{\Big((\sum^k_{n=0}\Omega_{n})^2+(\sum^k_{n=0}\Delta_{n})^2\Big)^{\frac{3}{2}}}\Big)
	\end{align}

	The control signals are depicted in figure\ref{2}($\mathrm{a_2,b_2}$) and the fidelity of gate optimization is depicted in figure\ref{2}($\mathrm{c_2}$). After 26 sample of 1 microsecond control pulses we reach to the fidelity of $0.998$ for beam-splitter gate optimization and the atomic beam-splitter gate at this optimal control reach to:
	\begin{align}
		U_{atomBS}(N\delta t)=\begin{pmatrix}
			0.704-0.07i&&0.707i\\
			0.707i&&0.704+0.07i
		\end{pmatrix} , N=26 , \delta t=1\mu sec , 0.998
	\end{align}
	
	By gate optimization, simultaneously we can control two populations of atoms, one population from state $|0\rangle$ to $|y+\rangle$ and other population from $|1\rangle$ to $|y-\rangle$ as we can see in figure\ref{2}($\mathrm{d_2}$).
		
	\appendix
	
	\bibliography{References}
\end{document}